\documentclass{ptptex}
\usepackage{graphicx}

\newcommand{\ket}[1]{|#1\rangle}
\newcommand{\braket}[2]{\langle #1|#2\rangle}
\newcommand{\mtrix}[3]{\langle \,#1\,|\,#2\,|\,#3\,\rangle}

\title{Sum Rules of the Multiple Giant Dipole States}

\author{Haruki \textsc{Kurasawa}$^{1}$, 
Toshio \textsc{Suzuki}$^{2}$ and
Pier Francesco \textsc{Bortignon}$^{3}$
}
\inst{
$^1$ Department of Physics, Faculty of Science, Chiba University, Chiba
263-8522, Japan \\

$^2$
Department of Applied Physics, Fukui University, Fukui 910-8507, Japan\\

$^3$
Dipartimento di Fisica, Universit$\grave{\rm a}$ di Milano, and INFN,\\
Sezione di Milano, Via Celoria 16, I-20133 Milano, Italy
}

\abst{
Various sum rules for multiple giant dipole resonance states
are derived. For the triple giant dipole resonance states, 
the energy-weighted sum of the transition strengths
requires a model to be related to those of the single and double giant
dipole resonance states.
It is also shown that the non-diagonal matrix elements of the double
commutator between the dipole operator and the nuclear Hamiltonian
give useful identities for the excitation energy
and transition strength of each excited state. 
Using those identities, the relationship between width of the single
dipole state and those of the multiple ones is qualitatively discussed.
}

\begin{document}
\notypesetlogo
\maketitle

It is well known that sum rules play an important role in a wide range of
physics\cite{suzuki}.
On the one hand, if an experiment shows breaking of the appropriate
sum rule, the basic framework of the model should be improved.
For example, 
the enhancement of the Thomas-Reiche-Kuhn(TRK) sum rule value
requires meson-exchange currents\cite{suzuki, bohr}.
The quenching of the Coulomb\cite{ks1}
and Ikeda\cite{ikeda} sum rule values indicates that nucleon-degrees of
freedom are not enough for understanding of the relevant nuclear
phenomena. 
On the other hand, when we
make some approximations in calculations, we should take care of the
fundamental sum rules. From this point of view,
the sum rules in the random phase approximation
have been explored by many authors in non-relativistic\cite{thouless} and
relativistic\cite{ks2} frameworks. 

In nuclear physics,
a typical example of the sum rules is obtained by the relationship
\begin{eqnarray}
[ D, [ H, D ]] = C,  \label{starting}
\end{eqnarray}
where $H, D,$ and $C$  denote the Hamiltonian of the system,
the excitation operator, and a c-number, respectively.
When we take the ground-state expectation value of the both sides, 
we have the model-independent sum rule,
\begin{eqnarray}
2\sum_n \omega_n |\mtrix{n}{D}{\,}|^2 = C,
\end{eqnarray}
where $\omega_n$ represents the excitation energy of the eigen state
of the Hamiltonian $\ket{n}$
from the ground state $\ket{\,\,}$.
For example, the TRK sum rule for dipole excitations, 
\begin{eqnarray}
\sum_n \omega_n |\mtrix{n}{D}{\,}|^2 = \frac{1}{2m}\frac{NZ}{A},\label{trk}
\end{eqnarray}
is obtained for
\begin{eqnarray}
D=\frac{Z}{A}\sum_{i=1}^N z_i- \frac{N}{A}\sum_{i=1}^Z z_i,
\end{eqnarray}
assuming its double commutator with the nuclear interactions to be
zero.
Recently in the same approximation as for the TRK sum rule, 
another sum rule has
been derived\cite{kurasawa},
\begin{eqnarray}
\sum_n\omega_n|\mtrix{n}{D^2}{\,}|^2 = 4S_0(1)S_1(1),\label{dgd}
\end{eqnarray}
where $S_1(1)$ stands for the TRK sum rule value and $S_0(1)$ is the
non-energy weighted sum of the dipole transition strengths,
\begin{eqnarray}
S_0(1)=\sum_n|\mtrix{n}{D}{\,}|^2.
\end{eqnarray}
The new sum rule Eq.(\ref{dgd}) is related to the double giant dipole
resonance(DGDR) states which have been recently observed through
relativistic heavy ion reactions\cite{bortignon}.
It implies that the energy-weighted sum of the transition
strengths for the DGDR is determined by the excitation energies and
strengths of the single giant dipole resonance(SGDR) states.  
This relationship is useful both for the analysis of experimental data
and for making models of the DGDR,
since the SGDR is well known. 

The purpose of the present paper is to derive other sum rules
for multiple dipole states
on the basis of the same assumption as for the TRK sum rule.
In particular, 
we will discuss the sum rules for the operator $D^3$, which is
 related to the triple giant dipole resonance(TGDR) states.
 We will also show that the non-diagonal matrix
 elements of Eq.(\ref{starting}) provide us with useful identities to
 understand the structure of the multiple giant resonance states.
 They give a constraint not only on the sum of the matrix elements,
 but also on each energy-weighted matrix element.


First we discuss sum rules for the operator $D^3$.
For convenience, let us define the notation
for the sum of the matrix elements,
\begin{eqnarray}
S_k(m)=\sum_n\omega_{mn}^k|\mtrix{mn}{D^m}{\,}|^2,
\end{eqnarray}
where $\omega_{mn}$ denotes the excitation energy of the $n$-th state
of the $m$-times SGDR, $\ket{mn}$. In this notation, the TRK sum
rule Eq.(\ref{trk})
and the new sum rule Eq.(\ref{dgd}) mentioned  above are expressed as
\begin{eqnarray}
S_1(1)&=&\sum_n \omega_{1n} |\mtrix{1n}{D}{\,}|^2,\\
S_1(2)&=&\sum_n\omega_{2n}|\mtrix{2n}{D^2}{\,}|^2.
\end{eqnarray}

The above two sum rules are calculated  with the use of the closure
property of the intermediate states.
For example, $S_1(2)$ is expressed as
\begin{eqnarray}
S_1(2)=\frac{1}{2}\left(\sum_n\mtrix{\,}{[ D^2, H ]}{2n}\mtrix{2n}{D^2}{\,} +
		   \mtrix{\,}{D^2}{2n}\mtrix{2n}{[ H, D^2 ]}{\,}\right).
\label{dgd2}
\end{eqnarray}
This equation is rewritten with the closure property as
\begin{eqnarray}
S_1(2)=\frac{1}{2}\mtrix{\,}{[ D^2, [ H, D^2 ]]}{\,}.
\end{eqnarray}
By calculating explicitly the double commutator,
we obtain Eq.(\ref{dgd}).

Unfortunately, for $S_1(3)$ we can not use the same procedure, because
the final states excited by $D^3$ are not only $\ket{3n}$, but also
$\ket{1n}$. In the case of Eq.(\ref{dgd2}),
$D^2$ has also a matrix element between the
ground states, but it does not contribute to the energy-weighted sum and
therefore, we could use the closure property.
In other words, the sum rule value obtained by the double
commutator of $D^3$ with Hamiltonian is not the value of $S_1(3)$.
Keeping this fact in mind, we calculate
\begin{eqnarray}
S_1(f)&=&\sum_f\omega_f|\mtrix{f}{D^3}{\,}|^2 \nonumber \\
 &=& \frac{1}{2}\mtrix{\,}{[D^3, [H, D^3]]}{\,}
 =9S_1(1)\left(S_0^2(1)+S_0(2)\right),\label{f}
\end{eqnarray}
where we have used
\begin{eqnarray}
\mtrix{\,}{D^4}{\,}=S_0^2(1)+S_0(2).
\end{eqnarray}
Thus the value of $S_1(f)$ is fixed by the TRK sum rule values and
the transition strengths of the SGDR and DGDR. 
In the above equation, however, $\ket{f}$
should be either $\ket{1n}$ or $\ket{3n}$.
Fig.~1 shows the transitions to be included in $S_1(f)$. In addition to (a)
corresponding to $S_1(3)$, four kinds of transitions contribute to
$S_1(f)$.

\begin{figure}
\centerline{\includegraphics{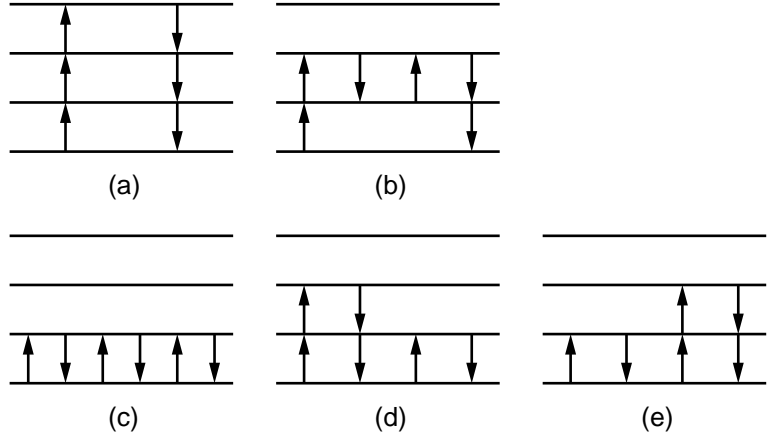}}
\caption{Transitions induced by $D^3$}
\label{fig1}
\end{figure}

\vspace{0.5\baselineskip}

Among the transitions in Fig.~1, the sums of the strengths in (c),
(d), and (e) are calculated individually.
The contribution (c)
is written as
\begin{eqnarray}
{\rm (c)}&=&\sum_{n',n'',n'''}\omega_{1n'}\mtrix{\,}{D}{1n}\mtrix{1n}{D}{\,}
\mtrix{\,}{D}{1n'}\mtrix{1n'}{D}{\,}\mtrix{\,}{D}{1n''}\mtrix{n''}{D}{\,}
\nonumber\\
 \noalign{\vskip4pt}
&=& S_1(1)S_0^2(1).\label{b}
\end{eqnarray}
The sums of (d) and (e), on the other hand, are described as
\begin{eqnarray} 
({\rm d})=({\rm e})&=&\sum_{n,n',n'',n'''}\omega_{1n''}
 \mtrix{\,}{D}{1n}\mtrix{1n}{D}{2n'}
 \mtrix{2n'}{D}{1n''}  \nonumber \\
& &
\phantom{\sum\ \ \omega\ \ } 
\times
\mtrix{1n''}{D}{\,}\mtrix{\,}{D}{1n'''}\mtrix{1n'''}{D}{\,}\nonumber\\
&=& \frac{1}{2}\mtrix{\,}{[D^3, [H,D]]}{\,}S_0(1)-S_1(1)S_0^2(1)
=2S_0^2(1)S_1(1).\label{de}
\end{eqnarray}
From Eqs.(\ref{f}), (\ref{b}) and (\ref{de}), 
we have the sum of (a) and (b)
\begin{eqnarray}
& &S_1(3)+\sum\omega_{1n''}\mtrix{\,}{D}{1n}\mtrix{1n}{D}{2n'}
\mtrix{2n'}{D}{1n''} \nonumber \\
& &
\phantom{S_1(3)+\sum}
\times
\mtrix{1n''}{D}{2n'''}\mtrix{2n'''}{D}{1n''''}\mtrix{1n''''}{D}{\,}\nonumber\\
 &=& 4S_1(1)S_0^2(1)+9S_1(1)S_0(2).\label{ac}
 \end{eqnarray}
Thus, although several sum rules are obtained for $D^3$,
we can not obtain the relationship of $S_1(3)$ to the
energy-weighted and non-energy-weighted sum for the strengths of the SGDR and
DGDR. The second term of the left hand side
in the above equation, coming from (b), can not be
expressed in terms of $S_k(m)$.

In order to separate the contribution $S_1(3)$ from (b),  we need an
assumption.
For example, if we assume the folding model where the DGDR states
are composed of the dipole bosons $\ket{1n}$,
\begin{eqnarray}
\ket{2n}=\ket{1n',1n''}\quad(n'\geq n''),
\end{eqnarray}
then we have
\begin{eqnarray}
({\rm b})=4S_1(1)S_0^2(1).\label{fm}
\end{eqnarray}
Here, we have used the equation of the folding model,
\begin{eqnarray}
\sum_{n''}\mtrix{1n}{D}{2n''}\mtrix{2n''}{D}{1n'}
=\delta_{nn'}S_0(1)+\mtrix{\,}{D}{1n'}\mtrix{1n}{D}{\,},\label{boson}
\end{eqnarray}
which is obtained from the fact that
\begin{eqnarray}
\mtrix{1n}{D}{1n',1n''}
 &=&\sqrt{2}\mtrix{\,}{D}{1n}\delta_{nn'}\delta_{n'n''}
+\mtrix{\,}{D}{1n''}\delta_{nn'}(1-\delta_{n'n''}) \nonumber \\
 \noalign{\vskip4pt}
& &
 +\,\mtrix{\,}{D}{1n'}\delta_{nn''}(1-\delta_{n'n''}). \label{boson2}
\end{eqnarray}
Eqs.(\ref{ac}) and (\ref{fm}) yield
\begin{eqnarray}
S_1(3) = 9S_1(1)S_0(2).
\end{eqnarray}

We note that Passos et al.\cite{passos}
have defined TGDR states which are excited by the operator
\begin{eqnarray}
O^{(3)}=D^3-D\frac{\mtrix{\,}{D^4}{\,}}{\mtrix{\,}{D^2}{\,}}.
\end{eqnarray}
This operator is determined with the requirement
that the doorway state of the TGDR is orthogonal to the one of the SGDR,
 \begin{eqnarray}
\sum_n\mtrix{\,}{D}{1n}\mtrix{1n}{O^{(3)}}{\,}=0.
\end{eqnarray}
Their TGDR states defined in this way are different from the present 
$\ket{3n}$ which satisfies $\braket{1n}{3n'}=0$.


The sum rules discussed so far are related to the diagonal matrix
elements of Eq.(\ref{starting}).
Next we will show that
if we calculate its non-diagonal matrix elements,
we can obtain other useful identities.

The equation
\begin{eqnarray}
\mtrix{m+1, n}{[D, [H, D]]}{m-1,n'}=0
\end{eqnarray}
provides us with 
\begin{eqnarray}
& &(\omega_{m+,n}+\omega_{m-1,n'})
 \sum_{n''}\mtrix{m+1,n}{D}{mn''}\mtrix{mn''}{D}{m-1,n'}\nonumber\\
&=&2\sum_{n''}\omega_{mn''}\mtrix{m+1,n}{D}{mn''}\mtrix{mn''}{D}{m-1,n'}.
 \label{general}
\end{eqnarray}
For m=1 and 2, for example, we have the relationships,
\begin{eqnarray}
\omega_{2n}\mtrix{2n}{D^2}{\,}
 &=&2\sum_{n'}\omega_{1n'}\mtrix{2n}{D}{1n'}
 \mtrix{1n'}{D}{\,},\label{2omega}\\
\omega_{3n}\mtrix{3n}{D^3}{\,}
 &=&3\sum_{n'}\omega_{1n'}\mtrix{3n}{D^2}{1n'}
 \mtrix{1n'}{D}{\,}.\label{3omega}
\end{eqnarray}
In deriving the last equation, we have used Eq.(\ref{2omega}),
together with Eq.(\ref{general}). 
According to these identities, it is possible to make the
following  comments.

First, the excitation energy $\omega_{2n}$ is determined by
$\omega_{1n}$ and transition strengths. This fact is also true for other
$\omega_{mn}$. The sum rule Eq.(\ref{dgd}) is the relationship between the
sum of the energy-weighted strengths for the DGDR and those of
the SGDR, while Eq.(\ref{2omega}) provides the relationship of each
energy-weighted strength of $\ket{2n}$ to $\omega_{1n}$ and transition
strengths,
\begin{eqnarray}
\omega_{2n}\mtrix{\,}{D^2}{2n}\mtrix{2n}{D^2}{\,}=2\sum_{n'}\omega_{1n'}
\mtrix{\,}{D^2}{2n} \mtrix{2n}{D}{1n'}\mtrix{1n'}{D}{\,}.\label{2omega2}
\end{eqnarray}

In using the above equation, the energy-weighted sum of the DGDR is
expressed in the form:
\begin{eqnarray}
S_1(2)=2\sum_{n,n'}\omega_{1n'}\mtrix{\,}{D^2}{2n}
 \mtrix{2n}{D}{1n'}\mtrix{1n'}{D}{\,}.
\end{eqnarray}
Since the right-hand side of this equation is expressed as
\begin{eqnarray}
{\rm r.h.s.} = 2\sum_n\omega_{1n}\mtrix{\,}{D^3}{1n}
 \mtrix{1n}{D}{\,}-2S_0(1)S_1(1),
\end{eqnarray}
we have another sum rule,
\begin{eqnarray}
\sum_n\omega_{1n}\mtrix{\,}{D^3}{1n}\mtrix{1n}{D}{\,}=3S_0(1)S_1(1).
\end{eqnarray}
Of course, it is  possible to derive the last sum rule by calculating the
double commutator, as was done in Eq.(\ref{de}).
In this case, the sum rule of the DGDR, Eq.(\ref{dgd}),
can be obtained from Eq.(\ref{2omega}). We note also that Eq.(\ref{de})
is easily obtained in employing Eq.(\ref{2omega2}).

Second, if there is a single collective state $\ket{m=1}$ with the excitation
energy $\omega_1$, then Eq.(\ref{2omega}) yields $\omega_{2n}=2\omega_1$,
and hence, owing to Eq.(\ref{general}), we have $\omega_{mn}=m\omega_1$.
More approximately, Eq.(\ref{2omega}) implies that if
$\omega_{1n}\approx \bar{\omega}$, we should have $\omega_{2n} \approx
2\bar{\omega}$, where $\bar{\omega}$ represents the mean energy of the SGDR.
This fact is expressed by rewriting Eq.(\ref{2omega}) as
\begin{eqnarray}
\omega_{2n}\mtrix{2n}{D^2}{\,}
 =2\sum_{n'}(\omega_{1n'}-\bar{\omega})\mtrix{2n}{D}{1n'}
 \mtrix{1n'}{D}{\,}+2\bar{\omega}\mtrix{2n}{D^2}{\,}.
\end{eqnarray}
Thus, if the width of the SGDR is narrow, the width of
the DGDR is also expected to be narrow.
The conclusion on other m-GDR is also the same. 
In other word, if the width of the SGDR is narrow, interactions satisfying
$[D, [V, D]] =0$ may not play a role to make the width of the m-GDR
broader.
On the one hand, the diagonal matrix element of Eq.(\ref{starting})
shows that there should be the same dipole strength on any state.
This fact is usually called Brink's hypothesis\cite{bortignon}.
On the other hand, its off-diagonal
matrix elements provide a constraint on the distribution of the strengths.

The constraint on the width of the DGDR due to the SGDR may be seen more
clearly  in the following identity
which is obtained from Eq.(\ref{2omega}),
\begin{eqnarray}
& &\sum_n(\omega_{2n}-2\bar{\omega})^2|\mtrix{2n}{D^2}{\,}|^2 \nonumber \\
&=&4\sum_{n,n'}(\omega_{1n}-\bar{\omega})(\omega_{1n'}-\bar{\omega})
 \mtrix{\,}{D}{1n}\mtrix{1n}{D^2}{1n'}\mtrix{1n'}{D}{\,},\label{md}
\end{eqnarray}
where the mean energy of the SGDR is defined by $\bar{\omega}=S_1(1)/S_0(1)$.

In the folding model, the above equation yields the relationship between
the variances of the DGDR and SGDR strength functions.
Eq.(\ref{boson}) gives 
\begin{eqnarray}
S_0(2)=2S_0^2(1),\label{noen}
\end{eqnarray}
and
\begin{eqnarray}
 \sum_n\omega_{2n}|\mtrix{2n}{D^2}{\,}|^2/S_0(2)=2\bar{\omega}. \label{2mean}
\end{eqnarray}
The straightforward calculation of Eq.(\ref{md}),
together with the above two equations and Eq.(\ref{boson}), provides us with
the well-known result\cite{kurasawa,bortignon},
\begin{eqnarray}
\sigma_2 =\sqrt{2}\sigma_1,\label{vdgd}
\end{eqnarray}
where $\sigma_1$ and $\sigma_2$ denote the variances of the SGDR and DGDR,
respectively,
\begin{eqnarray}
\sigma_1^2
 &=& \sum_n(\omega_{1n}-\bar{\omega})^2|\mtrix{1n}{D}{\,}|^2 / S_0(1),\\
\sigma_2^2
 &=& \sum_n(\omega_{2n}-2\bar{\omega})^2|\mtrix{2n}{D^2}{\,}|^2 / S_0(2).
\end{eqnarray}
It should be noted that the broadening by the factor $\sqrt{2}$
in Eq.(\ref{vdgd}) is not due to
nuclear interactions. This is just a stochastic factor
given for the two-dimensional probability distribution, when there is no
correlation between the two stochastic variables.
In the folding model, the stochastic variable is the excitation energies
of the bosons, and the probability function is related to the transition
strength function. Since the covariance in the two boson sets is zero in
the folding model, the only first term of the right
hand side in Eq.(\ref{boson}) contributes
to the right hand side of Eq.(\ref{md}).
Eqs.(\ref{noen}) and (\ref{2mean}) are also derived from a stochastic
point of view. 
In the folding model, we have $\sigma_m=\sqrt{m}\sigma_1$.

In summary, various sum rules for multiple giant dipole
states are derived by assuming, in the same way as for
the Thomas-Reiche-Kuhn sum rule, that the double commutator of
the dipole operator with the nuclear Hamiltonian is a constant.
Nuclear interactions depending only on nucleon coordinates,
exchange forces with zero-range and  spin-orbit forces
satisfy this assumption. It has been discussed that
the sum rule  for the triple giant dipole resonance 
can not be obtained without further assumptions.

The sum rules are derived from the diagonal
matrix elements of the double commutator. We have shown that
non-diagonal matrix elements also provide us with useful identities,
which yield constraints on the excitation energy and transition
strength of each nuclear state.
By using those identities, we can discuss qualitatively
the relationship between the width of the single giant dipole
resonance and those of the multiple ones. If the width of the single
dipole giant resonance is narrow, then that of the multiple one
is expected to be also narrow.
For more quantitative understanding of excited states, and
effects of exchange forces with finite range or velocity-dependent forces,
of course, more elaborate calculations are required.

Finally, we note that 
in fact, throughout this paper, we did not use the explicit form of the
dipole operator, but assumed only
the double commutator of the excitation operator
with the Hamiltonian to be constant. 
Therefore, the present study may be useful not only in nuclear physics,
but also in discussions of other quantum systems like atomic clusters and
Bose-Einstein condensation.
Moreover, collective excitations other than the dipole one may be
discussed in the same way. The new identities, derived from the
non-diagonal matrix elements of the double commutator,
may be useful for discussions on
anharmonicity of collective motions in many-body systems.

\end{document}